\documentclass[aps,prl,twocolumn,showpacs,groupedaddress,superscriptaddress,amsmath,amssymb]{revtex4}
\usepackage{graphicx}
\usepackage{natbib}

\begin{document}

\newcommand{\Rb}{{${}^{87}$Rb}}

\title{Slow light with integrated gain and large pulse delay}

\author{Irina Novikova}
\affiliation{Harvard-Smithsonian Center for Astrophysics, Cambridge,
Massachusetts, 02138}\affiliation{Department of Physics, College of
William\&Mary, Williamsburg, Virginia 23185, USA}
\author{David F. Phillips}
     \affiliation{Harvard-Smithsonian Center for Astrophysics,
Cambridge, Massachusetts, 02138}
\author{Ronald L. Walsworth}
\affiliation{Harvard-Smithsonian Center for Astrophysics, Cambridge,
Massachusetts, 02138} \affiliation{Department of Physics, Harvard
University, Cambridge, Massachusetts, 02138}

\date{\today}

\begin{abstract}
   We demonstrate slow and stored light in Rb vapor with a combination
of desirable features: minimal loss and distortion of the pulse shape, and
large fractional delay ($> 10$). This behavior is enabled by: (i) a group index
that can be controllably varied during light pulse propagation; and (ii)
controllable gain integrated into the medium to compensate for pulse loss. Any
medium with the above two characteristics should be able to realize similarly
high-performance slow light.
\end{abstract}


%
%

\maketitle


Optical buffers with controllable delay are key components for both
photonic optical networks and quantum information
processing systems. Requirements for such optical buffers are
an adjustable delay time (i.e., group index) for input signal pulses
over a wide range of
bandwidths, minimal pulse distortion and loss~\cite{tuckerJLT05}, and high
compression of the input pulse for high data density inside the
delay medium.
Large pulse delay (``slow light'') is achievable in many media including gas
vapors~\cite{lukin03rmp,boydPO,bigelowScience03,okawachiOE06,howellPRA06},
cold atoms~\cite{hau}, doped crystals~\cite{manson05}, photonic
bandgap crystals~\cite{mori07}, semiconductor
heterostructures~\cite{ku}, microresonators~\cite{totsuka07}, and
optical fibers~\cite{patnaik,linJOSAB04,mok,kaloshaPRA07}. In these
systems the group index is controllable using
a variety of techniques including the application of a strong optical control
field, varying the density of coupled atoms, or the coupling of a
microresonator to a waveguide.
However, large delay-bandwidth products are very challenging to obtain
because residual absorption of the input signal pulse typically
increases exponentially with the length
of the medium.
Here, we demonstrate a technique that provides independent control of
the signal pulse group velocity and amplitude, using (i) a temporally varying
group index in coordination with (ii) integrated gain in the medium.
This combination of features allows for
large fractional pulse delay (delay-bandwidth product $\gg 1$) with
minimal distortion and absorption of the
output pulse. The technique is general: any system with the two key
characteristics of controllable group index and integrated gain
should be able to realize high-performance slow light. We note that
excellent progress toward this goal has also been made recently with
a distinct technique involving gap solitons in an optical fiber Bragg
grating~\cite{mok}.
\begin{figure}
\includegraphics[width=1.00\columnwidth]{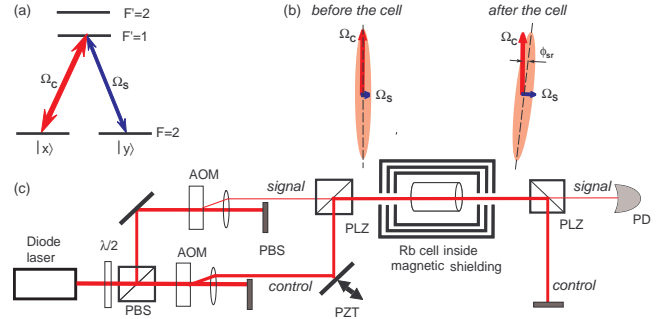}
      \caption{
(a) Simplified $\Lambda$-scheme for ${}^{87}$Rb atoms interacting with
degenerate, co-propagating control and signal fields with orthogonal linearly
polarizations. $|x\rangle$ and $|y\rangle$ represent appropriate superpositions
of magnetic ($m_F$) sublevels. (b) Representation of the elliptical
polarization of the total optical field created by the sum of the control and
signal fields with a relative phase difference $\theta$; and polarization
self-rotation of this total optical field by angle $\phi_{sr}$ after passing
through the EIT medium, which provides integrated gain for the signal field.
(c) Schematic of the experimental setup (see the text for abbreviations).
      \label{setup.fig}
}
\end{figure}

The steep dispersion necessary to create a large group index is
usually achieved in a very narrow frequency band through a resonant
transmission
feature. The achievable pulse delay-bandwidth product is therefore
limited ($\lesssim 1$)
by pulse absorption in the frequency wings of the
resonance since losses grow exponentially as a high-bandwidth pulse
propagates through a
high group-index (narrow transmission bandwidth)
medium~\cite{boydPRA05}. Increasing the bandwidth of the resonant
medium typically reduces the group index proportionally, such that
the delay-bandwidth product remains limited. However, a medium with
integrated
gain can compensate for absorption and
allow large pulse delays to be obtained.
Even in the presence of gain, the finite bandwidth of the resonance
leads to pulse distortion.  Nevertheless, as we show here, large
pulse delays with minimal
distortion and attenuation can be created through the combined use of
integrated gain and
dynamic control of the group index and hence the instantaneous group
velocity inside the medium.

Our demonstration experiments employed a dynamic form of
electromagnetically induced transparency (EIT) in warm Rb
vapor~\cite{lukin03rmp,fleischhauer00,phillips,mair,fleischhauer02}.
In EIT, a strong
control field determines the propagation of a weak signal pulse
interacting near resonance with an ensemble of radiators such as
atoms, typically in a $\Lambda$-scheme (see Fig.~\ref{setup.fig}(a)).
The
group velocity $v_g$ of the signal is
\begin{equation}
    v_g =  \frac{c}{1+\eta \, N / |\Omega_C|^2} =  \frac{c}{1+n_g},
\label{e.group}
\end{equation}
where $\Omega_C$ is the Rabi frequency induced by the control field on
the relevant atomic transition, $N$ is the atomic number density, and
$\eta = 3/(4\pi)\lambda^2c\gamma$ is the coupling constant between the
signal field and the atomic transition with $\lambda$ the optical
wavelength, $c$ the vacuum speed of light, and $\gamma$ the optical
decoherence rate. The group index, $n_g = \eta \, N / |\Omega_C|^2$. In
many EIT media, slow light ($v_g \ll c$) has been
demonstrated~\cite{lukin03rmp,boydPO,bigelowScience03,okawachiOE06,howellPRA06,manson05}.
The group delay,  $\tau_g$, of a signal pulse in such an EIT medium
(relative to
propagation in vacuum) is given by
\begin{equation}
\label{e.delay}
\tau_g \approx L/v_g \approx d \gamma / |\Omega_C|^2
\end{equation}
where $L$ is the length of the atomic medium and $d=\eta N L / (c
\gamma)$ is the optical depth.  The bandwidth of the EIT resonance,
$\gamma_{\mathrm{EIT}}$, for the propagating pulse through the dense
atomic medium is
\begin{equation}
\gamma_{\mathrm{EIT}}\approx \frac{|\Omega_C|^2}{\gamma \sqrt{d}}.
\label{e.bandwidth}
\end{equation}
Residual signal-field absorption
in an EIT medium,
$\alpha=\ln\left(P_{\mathrm{out}}/P_{\mathrm{in}}\right)$, is
well-approximated by~\cite{lukin03rmp}
\begin{equation}
\alpha\approx d \,  \frac{\gamma_0 \gamma}{|\Omega_C|^2}
\label{e.absorb}
\end{equation}
where $\gamma_0$ is the ground-state decoherence rate of the atomic system.

\begin{figure}
\includegraphics[width=0.80\columnwidth]{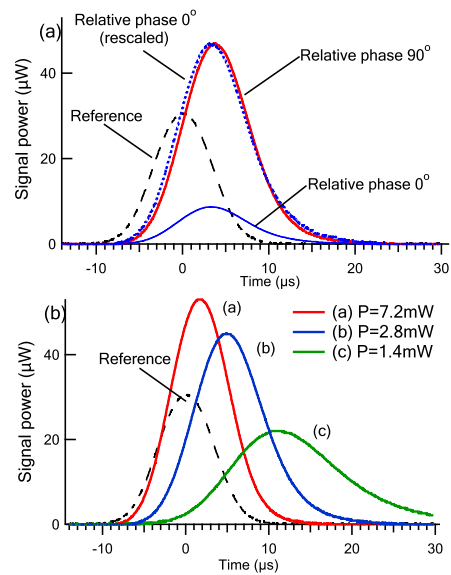}
\caption{(a) Measured output signal pulses for relative phase $\theta=0^\circ$
and $90^\circ$ between the control and signal fields, illustrating that
integrated gain provided by polarization self-rotation in an EIT medium affects
pulse amplitude but not pulse delay or shape. A rescaled signal pulse for
$\theta=0^\circ$ is also shown for easy comparison of the pulse shapes. The
control field power is constant at 3.5 mW for all traces. (b) Measured output
signal pulses, amplified by polarization self-rotation with $\theta=90^\circ$,
for different (constant) control field powers. These measurements illustrate
that a signal pulse amplified by integrated gain retains the usual EIT behavior
of pulse delay and bandwidth scaling inversely and linearly with the control
field intensity, respectively. Measurements were performed using a ${}^{87}$Rb
vapor cell filled with 40 Torr Ne buffer gas and heated to $72^\circ$C. A
reference input signal pulse is shown in each figure for comparison.
      \label{slow_flat.fig}
    }
\end{figure}

Significant delay of a signal pulse while simultaneously preserving its
amplitude and shape requires both small group velocity (hence small
control field intensity) and large spectral bandwidth and low
absorption (hence large control field intensity). These two competing
conditions offset each other~\cite{boydPRA05}, such that the
delay-bandwidth product for EIT-based slow light ($\approx
\tau_g\gamma_{\mathrm{EIT}}$) is independent of the control field
intensity, $|\Omega_C|^2$, but proportional to the optical depth.
Therefore, a larger delay-bandwidth product requires a larger optical
depth, which comes
at the expense of exponentially greater residual absorption (even for
high-quality EIT). This increased absorption can be
compensated for with integrated gain, but can still lead to pulse shape
broadening and distortion due to fractionally larger absorption and
nonlinear dispersion in the wings of the transmission
resonance. As we show here, such pulse shape corruption can be
corrected by increasing the control field intensity (i.e., decreasing
the group index) as
the signal pulse enters the EIT medium. With this technique, the
leading edge of the pulse enters the medium with
the control field at low intensity (i.e., small $v_g$); whereas the
trailing edge
enters the medium with a stronger control field (i.e., larger $v_g$).
Thus, the trailing edge of the signal pulse has a smaller net delay
inside the atomic medium than the leading edge, which compresses the
temporal extent of the pulse. Tailored use of such
compression allows a signal pulse to propagate through a medium with a lower
mean group velocity (i.e., larger net delay) than would be possible
with a constant control field
while preserving the temporal pulse length and thus bandwidth.
Related theoretical proposals for
temporal pulse shape manipulation have recently been
made for a variety of slow light
media~\cite{buffaPRA04,arkhipkinPRA06,kaloshaPRA07}.

\begin{figure*}
\includegraphics[width=1.30\columnwidth]{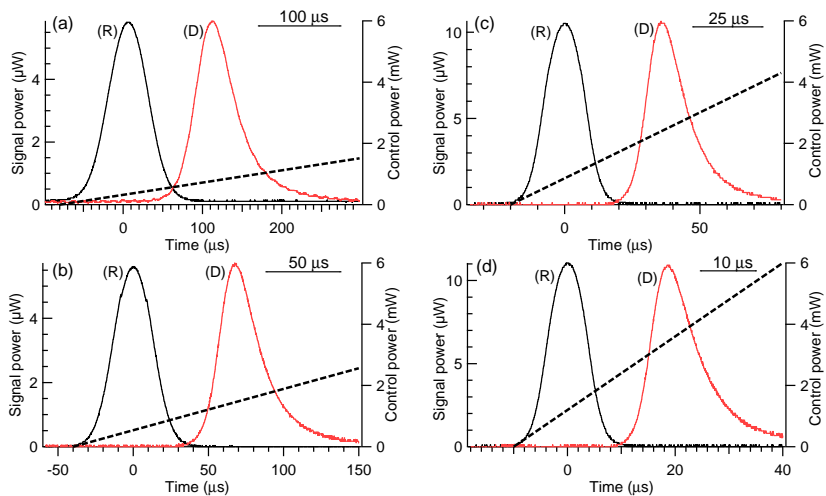}
\caption{Examples of measured signal pulse delay using a linearly increasing
control field intensity (i.e., decreasing group index) and integrated gain from
polarization self-rotation, for four widely-varying signal pulse bandwidths. In
each case the fractional pulse delay has been adjusted to $\approx 2$ and the
gain set for no net pulse attenuation.  (R) and (D) label the reference and
delayed signal pulses, correspondingly. These measurements were performed using
a ${}^{87}$Rb vapor cell filled with 25 Torr Ne buffer gas and heated to
$70^\circ$C.
      \label{slow_pulses.fig}
    }
\end{figure*}

The evolution of a signal pulse of amplitude $\Omega_S(z,t)$ as it
propagates through an EIT medium can be described by a simple
equation in the limit that $|\Omega_S|^2 \ll |\Omega_C|^2$ and
$\Omega_S$ varies
adiabatically~\cite{fleischhauer02}. In this limit, the internal
degrees of freedom of the atomic system can be evaluated
perturbatively and then eliminated, leaving the following propagation
equation for
the signal pulse amplitude near EIT resonance:
\begin{eqnarray}
\label{e.propagate}
\nonumber \left(\partial_t + c\partial_z\right)\Omega_S(z,t)=&&-\frac{\eta
N}{\Omega_C^*(t)}\left(\partial_t + \gamma_0 \right)
     \left( \frac{\Omega_S(z,t)}{\Omega_C(t)}\right) \\
& &+ \kappa_G \Omega_S(z,t).
\end{eqnarray}
Here $\kappa_G$ is the signal field gain associated with the medium (see
discussion below), $z$ is the propagation direction through the medium, and the
remaining parameters have been defined above.
In the absence of gain ($\kappa_G=0$) and with a constant control field
($\Omega_C(t)=\Omega_0$), Eq.~(\ref{e.propagate}) reverts to a simple
wave equation for
the signal field with a group velocity given by Eq.~(\ref{e.group}) and
absorptive losses by Eq.~(\ref{e.absorb}).  With appropriate ramping
of $\Omega_C$ and integrated gain in the medium (non-zero
$\kappa_G$), Eq.~(\ref{e.propagate}) qualitatively describes signal
pulse propagation with large fractional delay and insignificant loss
and distortion --- consistent with the experimental demonstrations
reported here.

In our experiments, we employed a degenerate $\Lambda$-system based on
ground-state Zeeman sublevels of ${}^{87}$Rb, with control and signal fields
having orthogonal linear polarizations (see Fig.~\ref{setup.fig}(a)). This
operational configuration provides EIT with integrated gain for the signal
field due to polarization
self-rotation~\cite{self-rot,novikovaOL00,novikovaJMO02}. In this well-studied
phenomenon, the major polarization axis of elliptically polarized light rotates
during propagation (see Fig.~\ref{setup.fig}(b)) due to a differential
refractive index for the light's two circularly-polarized components, induced
by an ac-Stark shift of the atomic Zeeman sublevels arising from off-resonant
excited states~\cite{SRnote}.
For a simplified four-level system the angle of polarization self-rotation
is~\cite{novikovaJMO02}:
\begin{equation} \label{phi_sr}
\phi_{sr} \simeq \frac{d \gamma }{2 \Delta_{HF}} \epsilon
= G \epsilon L,
\end{equation}
where $\Delta_{HF}$ is the effective detuning of the off-resonant
excited state~\cite{novikovaOL00}, $\epsilon$ is the ellipticity of
the total optical field, and we define $G=d\gamma / (2L \Delta_{HF})$
as the self-rotation coefficient.
A linearly-polarized field (such as the control field or signal field
alone) has zero ellipticity and thus suffers no rotation of its
polarization. However, when the orthogonally polarized control and
signal fields are both present, the total field is in general
elliptically polarized depending on the phase difference, $\theta$,
between the control and signal fields. The resulting polarization
self-rotation of the total field depends on the degree of ellipticity,
$\epsilon = (\Omega_S/\Omega_C)\sin{\theta}$.  The rotation of the
polarization ellipse serves to transfer a fraction of the control
field intensity into the signal field, leading to integrated gain of
the signal field, $\kappa_G \approx c \, G \sin{\theta}$.
Note, however, that this change in the signal field intensity does
not affect the
coherent properties of the medium; nor does it significantly decrease the
control field intensity (since $\phi_{sr} \ll 1$). Thus by varying the
polarization phase difference between the control and signal fields,
we can controllably vary the intensity of the signal pulse
significantly while having
little effect on the dynamics of signal pulse propagation.

We used the experimental setup shown schematically in
Fig.~\ref{setup.fig}(c) to
demonstrate this technique. An extended cavity diode laser was tuned
to 795~nm at the $F=2\rightarrow F^\prime=1$ $D_1$ transition of
${}^{87}$Rb.
Orthogonally polarized control and signal beams were
created by separating two
polarizations on a polarizing beam-splitter (PBS), sending them
through two separate phase-locked acousto-optical modulators (AOM), and then
recombining the first-order beams on a high-quality polarizer (PLZ)
with extinction ratio of $5\cdot 10^{-5}$. One of the mirrors was mounted on
a piezo-ceramic drive (PZT) which allowed the relative phase $\theta$ between
the control and signal fields to be adjusted by changing the path
length for the control field. Maximum total laser power at the Rb vapor cell
was $8$~mW, collimated into a $5$-mm beam diameter.
The ${}^{87}$Rb vapor cell was housed inside four
layers of magnetic shielding and heated conductively by a
blown-hot-air oven.  Two different cylindrical vapor cells were used
for the various measurements: each had length of 75~mm
and diameter of 25~mm and was filled with isotopically enriched
${}^{87}$Rb; one cell had 22~Torr and the other 40~Torr of Ne buffer
gas at room temperature. After the laser fields traversed the
cell, the signal field was filtered from the control field using a
high-quality polarizer (PLZ) and its intensity measured using a
photodetector (PD).

To optimize the signal-field gain, we adjusted the relative phase
$\theta$ between the control and signal fields by varying the
position of the mirror in
the control-field channel using the PZT, as shown in
Fig.~\ref{setup.fig}(c).  As expected, we found that changing
the relative phase $\theta$
between the control and signal fields from $0^\circ$ to $90^\circ$
changes the amplitude of the
signal pulse from its minimum (no self-rotation) to maximum, without
affecting the output signal field pulse shape or delay (see
Fig.~\ref{slow_flat.fig}(a)).  We also found that the dependence of
the delay of the
amplified signal pulse on control field intensity follows
Eq.~(\ref{e.delay}), as shown in Fig.~\ref{slow_flat.fig}(b): i.e., the delay
is inversely proportional to the control field
intensity. These measurements confirm that polarization self-rotation
acts as a form of integrated gain in the medium: it can increase the
signal pulse amplitude and
compensate for loss mechanisms, but does not affect the pulse delay or shape.
Measured output signal pulses corresponding to weaker control fields in
Fig.~\ref{slow_flat.fig}(b) are temporally broadened, consistent with
Eq.~\ref{e.bandwidth}.
We next showed that such temporal broadening of the signal pulse due
to finite EIT bandwidth can be
eliminated by smoothly increasing the control field intensity (i.e.,
decreasing the group index) as the signal pulse enters and traverses
the atomic medium.
Changing the control field intensity, which is uniform across the
length of the cell, creates a differential group delay for the front
and the back of the pulse leading to pulse compression which
compenates for broadening due to finite EIT bandwidth.
Fig.~\ref{slow_pulses.fig} shows several measured examples of the
combined use of linearly increasing control field intensity and
polarization self-rotation, which allows signal pulse shape- and
amplitude-preserving propagation with large fractional delay ($\approx
2$ in the examples shown) for a wide range of pulse bandwidths.

\begin{figure}
\includegraphics[width=1.00\columnwidth]{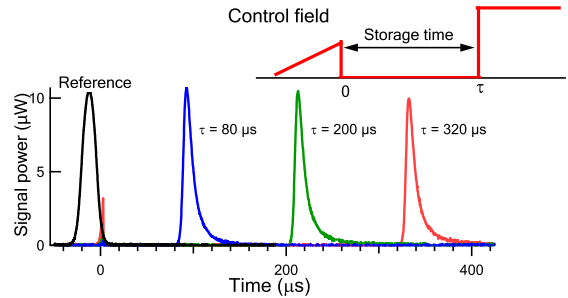}
\caption{
      \label{stored_pulses.fig}
      Examples of measured signal pulse delay using stored light with
a linearly-increasing control field intensity and integrated gain
from polarization self-rotation during the entry of the pulse into
the medium. The inset shows the timing of the control field.
Fractional pulse delay up to $\simeq 20$ is shown here with only
modest loss and pulse distortion. These measurements were performed
using a ${}^{87}$Rb vapor cell filled with 25 Torr Ne buffer gas and
heated to $70^\circ$C. A reference input signal pulse is shown for
comparison.}
\end{figure}

The combination of a dynamic control field and polarization
self-rotation can also be applied to stored
light~\cite{lukin03rmp,fleischhauer00,phillips,mair,fleischhauer02},
enabling very large fractional pulse delays as shown in
Fig.~\ref{stored_pulses.fig}. Once a slow-light signal pulse is
localized inside
the EIT medium, it can be mapped into a stationary collective spin
state by adiabatically switching off the control field, and later
mapped back into a propagating slow-light pulse by switching the
control field back on. Combining this stored light technique with a
linear ramp of the control field during entry of the signal pulse
into the Rb vapor, as well as integrated gain from polarization
self-rotation, we straightforwardly achieved fractional pulse delays
$> 10$ with minimal loss and distortion. Further optimization should
be possible by customizing the control field
temporal profile~\cite{arkhipkinPRA06}.

In summary, we demonstrated slow and stored light in Rb vapor with
minimal loss and pulse distortion and large fractional delay. This behavior is
enabled by the use of a medium with (i) a group index that can be
controllably varied during light pulse
propagation, which allows for large pulse delay and corrects for
distortion; and (ii) integrated, independently-controllable gain to
offset residual loss. The technique is general and should be
applicable to other atomic and solid-state systems, since a
controllable group index and gain
are common in many materials, including doped
crystals~\cite{manson05}, semiconductor heterostructures~\cite{ku},
and optical fibers~\cite{patnaik,linJOSAB04,mok,kaloshaPRA07}.

We are grateful to A.V. Gorshkov, Y. Xiao and M. Klein for useful discussions.
This work was supported by ONR, DARPA, NSF, and the Smithsonian Institution.




\begin{thebibliography}{99}
\bibitem{SRnote} Polarization self-rotation neither relies on
ground-state coherence
associated with EIT nor affects it and is also present in the absence
of any applied magnetic field.
\end{thebibliography}
\end{document}